\documentclass[superscriptaddress,reprint,amsmath,amssymb,aps,floatfix]{revtex4-1}

\usepackage{amsmath,gensymb,textcomp,bm,dcolumn,eurosym,array,tabu,multirow,nicefrac,color,subfigure,graphicx,upgreek}
\usepackage[colorlinks, linkcolor=blue, citecolor=blue, urlcolor=blue, breaklinks=true]{hyperref}

\usepackage{mathpazo,times} 

\newcommand{\bra}[1]{\langle #1 \rvert}
\newcommand{\ket}[1]{\lvert #1 \rangle}

\newcommand{\ketbra}[2]{\ket{#1}\bra{#2}}

\usepackage[subsectionbib]{bibunits}
\usepackage{graphicx}
\usepackage{dcolumn}
\usepackage{bm,MnSymbol}
\usepackage{units}
\usepackage{psfrag}
\usepackage{float}
\usepackage{color}

\begin{document}
\title{Verification of high-dimensional entanglement generated in quantum interference}



\author{Yuanyuan Chen}
\affiliation{
Institute for Quantum Optics and Quantum Information - Vienna (IQOQI), Austrian Academy of Sciences, Boltzmanngasse 3, 1090 Vienna, Austria.}
\affiliation{Vienna Center for Quantum Science \& Technology (VCQ), Faculty of Physics, University of Vienna, Boltzmanngasse 5, 1090 Vienna, Austria}
\affiliation{Department of Physics, Jiujiang Research Institute and Collaborative Innovation Center for Optoelectronic Semiconductors and Efficient Devices, Xiamen University, Xiamen 361005, China}

\author{Sebastian Ecker}
\affiliation{
Institute for Quantum Optics and Quantum Information - Vienna (IQOQI), Austrian Academy of Sciences, Boltzmanngasse 3, 1090 Vienna, Austria.}
\affiliation{Vienna Center for Quantum Science \& Technology (VCQ), Faculty of Physics, University of Vienna, Boltzmanngasse 5, 1090 Vienna, Austria}

\author{Jessica Bavaresco}
\affiliation{
Institute for Quantum Optics and Quantum Information - Vienna (IQOQI), Austrian Academy of Sciences, Boltzmanngasse 3, 1090 Vienna, Austria.}

\author{Thomas Scheidl}
\affiliation{
Institute for Quantum Optics and Quantum Information - Vienna (IQOQI), Austrian Academy of Sciences, Boltzmanngasse 3, 1090 Vienna, Austria.}
\affiliation{Vienna Center for Quantum Science \& Technology (VCQ), Faculty of Physics, University of Vienna, Boltzmanngasse 5, 1090 Vienna, Austria}

\author{Lixiang Chen}
\affiliation{Department of Physics, Jiujiang Research Institute and Collaborative Innovation Center for Optoelectronic Semiconductors and Efficient Devices, Xiamen University, Xiamen 361005, China}

\author{Fabian Steinlechner}
\email{Fabian.Steinlechner@iof.fraunhofer.de}
\affiliation{Fraunhofer Institute for Applied Optics and Precision Engineering IOF, Albert-Einstein-Strasse 7, 07745 Jena, Germany.}
\affiliation{Friedrich Schiller University Jena, Abbe Center of Photonics, Albert-Einstein-Str. 6, 07745 Jena, Germany.}

\author{Marcus Huber}
\affiliation{
Institute for Quantum Optics and Quantum Information - Vienna (IQOQI), Austrian Academy of Sciences, Boltzmanngasse 3, 1090 Vienna, Austria.}

\author{Rupert Ursin}
\email{Rupert.Ursin@oeaw.ac.at}
\affiliation{
Institute for Quantum Optics and Quantum Information - Vienna (IQOQI), Austrian Academy of Sciences, Boltzmanngasse 3, 1090 Vienna, Austria.}
\affiliation{Vienna Center for Quantum Science \& Technology (VCQ), Faculty of Physics, University of Vienna, Boltzmanngasse 5, 1090 Vienna, Austria}


\begin{abstract}
Entanglement and quantum interference are key ingredients in a variety of quantum information processing tasks. Harnessing the generation and characterization of entanglement in high-dimensional state spaces is a necessary prerequisite towards practical quantum protocols. Here, we use quantum interference on a beam splitter to engineer hyperentanglement in polarization and discrete frequency degrees of freedom (DOF). We show how independent measurements of polarization and frequency DOF allow for the verification of high-dimensional entanglement in the combined state space. These results may indicate new paths towards practical exploitation of entanglement stored in multiple degrees of freedom, in particular in the context of high-dimensional quantum information processing protocols.
\end{abstract}


\maketitle

\section{Introduction}
Quantum entanglement of photons is a crucial resource for quantum information applications such as quantum key distribution and quantum teleportation, as well as for studying fundamental physics in Bell experiments \cite{poppe2004practical,wang2015quantum,giustina2013bell,ma2009experimental}. Several degrees of freedom (DOF) of photons can be utilized to encode quantum entanglement, including polarization \cite{yin2017satellite}, spatial path \cite{rossi2009multipath}, orbital angular momentum \cite{fickler2012quantum}, time-bin \cite{halder2007entangling} and frequency \cite{ramelow2009discrete}. Some of these properties can exist independently of each other, which enables the entanglement of more than one property simultaneously, known as hyperentanglement \cite{barreiro2005generation,kwiat1997hyper,vergyris2019fibre}. Photon pairs entangled in multiple properties can carry more quantum information, making them compelling for high-capacity quantum communications. Encoding information in multiple degrees of freedom may also facilitate the implementation of certain quantum communication primitives: For instance, complete Bell state measurements can be performed deterministically for superdense coding or larger quantum states can be transmitted in quantum teleportation, thus increasing the capacity of classical and quantum channels \cite{barreiro2008beating,schuck2006complete,wang2015quantum}. Moreover, hyperentanglement can enhance the fidelity of mixed entangled states in entanglement purification and increase the state space for multi-photon entanglement and quantum computing \cite{sheng2010deterministic,xu2012demonstration,gao2010experimental,kok2007linear}.

High-dimensional quantum information processing has highlighted the need of verifying, certifying and quantifying the high dimensionality of hyperentanglement. The full determination of quantifying the amount of entanglement in high-dimensional quantum states is a daunting challenge, since the requirement of measuring a complete set of observables in a global state space is exponentially complex. Hence, it is of great significance to design wieldy and practical strategies to verify the amount of entanglement and its dimensionality, in particular with as few assumptions on the hyperentangled state as possible.

The objective of this work is twofold: First, we demonstrate how multi-photon interference on a beam splitter may itself be harnessed as a tool to engineer hyperentangled states. In our recent work \cite{chen2018polarization}, we utilized time-reversed Hong-Ou-Mandel (HOM) interference to generate polarization entanglement in two spatial modes without the usual requirement for distinguishability in an auxiliary degree of freedom. Here, we extend this approach to the generation of  hyperentanglement in polarization and discrete frequency modes.
Secondly, after characterizing the polarization and frequency interference for this state, we show how independent measurements performed on each of these degrees of freedom suffice to verify high-dimensional entanglement under minimal assumptions on the preparation of the state.

We believe that these results, demonstrating a path towards generating unconventional quantum states via quantum interference, as well as a practical way of extending results obtained for single degrees of freedom to the combined state space may prove valuable tools towards practical high-dimensional quantum information processing.

\section{Generation of hyperentanglement by Hong-Ou-Mandel interference}
\begin{figure}[!t]
\centering
\includegraphics[width=\linewidth]{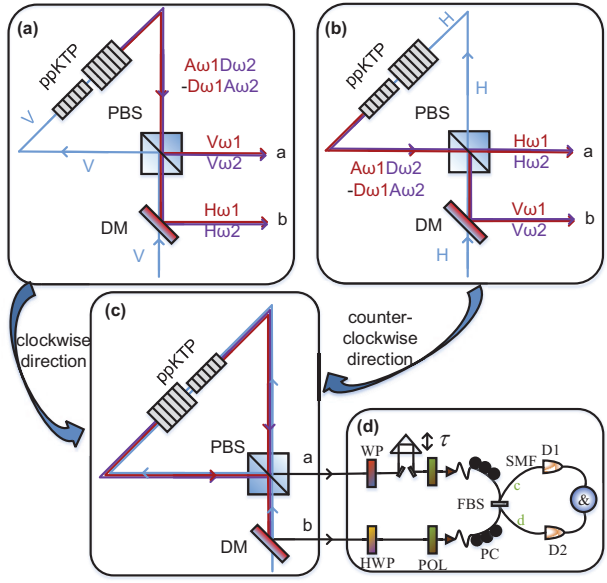}
\caption{Schematic of experimental setup. (a) Clockwise and (b) anti-clockwise directions of (c) Sagnac interferometer. (d) Hong-Ou-Mandel interferometer for frequency entanglement. DM: dichroic mirror, PBS: polarizing beam splitter, HWP: half wave plate, ppKTP: periodically poled potassium titanyl phosphate, POL: polarizer, $D_{1/2}$: single photon counting modulator, SMF: single mode fiber, PC: polarization controller, FBS: in-fiber beam splitter, WP: wave plate, a/b: output port of PBS, $c/d$: output port of FBS.}
\label{figure1}
\end{figure}
Entanglement can be engineered in a variety of physical systems \cite{barz2010heralded,delteil2016generation,riedel2012atom}, with spontaneous parametric down-conversion (SPDC) in nonlinear materials representing one of the most efficient ways reported to date. In the SPDC process, pump photons spontaneously decay into signal and idler photons, with conservation of momentum and energy resulting in entanglement of spatio-temporal properties.The generation, manipulation and detection of polarization-entangled \cite{steinlechner2017distribution,ursin2007entanglement,kaiser2012high} or frequency-entangled \cite{xavier2008full,peters2009dense} photons have already been extensively investigated and widely applied. Nevertheless, the manipulation is even trickier if the frequency-entangled photon pair can not be separated into two spatial modes. A discrete frequency-entangled Bell state can be represented as $\ket{\Psi^-_\omega}=\frac{1}{\sqrt{2}}(\ket{\omega_1}\ket{\omega_2}-\ket{\omega_2}\ket{\omega_1})$, where $\ket{\omega_{1/2}}$ are well-seperated single photon frequency bins. First approaches for generating this state relied on the projection of continuous frequency spectrum onto well-defined frequency bins prior to detection \cite{rarity1990two,ou1988observation}. A great number of schemes are proposed to create discrete frequency entanglement by using nonlinear waveguides \cite{ravaro2005nonlinear}, in-fiber Sagnac loops \cite{li2009all} and entanglement-transfer from the polarization domain \cite{ramelow2009discrete}. All of these schemes mainly focus on the generation of entanglement in the frequency domain, whereas simultaneous entanglement in other degrees of freedom would enable various hyperentanglement-assisted quantum information processing protocols.

Here, we present a polarization and discrete frequency hyperentanglement source by quantum interference. The key part of our source (see Fig.\ \ref{figure1}) consists of two periodically poled potassium titanyl phosphate (ppKTP) crystals designed for type-II quasi-phase matching. They are arranged in sequence and oriented with a relative inclination of 90$^\circ$ along their common propagation axis. These crossed crystals are placed at the center of a polarization Sagnac interferometer, which is bi-directionally pumped with a continuous wave laser. In the clockwise direction of the Sagnac interferometer (see Fig.~\hyperref[figure1]{\ref*{figure1}(a)}), the pump laser can either create a photon pair in the first crystal $\ket{A_{\omega_1}}\ket{D_{\omega_2}}$ or in the second crystal $\ket{D_{\omega_1}}\ket{A_{\omega_2}}$. Since both events occur with equal probability, the resultant quantum state reads
\begin{equation}
\begin{split}
\ket{\psi}_{\text{CW}}=\frac{1}{\sqrt{2}}(\ket{A_{\omega_1}}\ket{D_{\omega_2}}+e^{i\phi}\ket{D_{\omega_1}}\ket{A_{\omega_2}}),
\end{split}
\label{eq:clockwiseAD}
\end{equation}
where (A)D denotes (anti-)diagonal polarization, $\omega_{1}$ and $\omega_{2}$ are two well-separated frequency bins and $\phi$ is the relative phase factor. By setting $\phi=\pi$, the state can be rewritten in the H/V polarization basis as
\begin{equation}
\begin{split}
\ket{\psi}_{\text{CW}}=\frac{1}{\sqrt{2}}(\ket{H_{\omega_1}}\ket{V_{\omega_2}}-\ket{V_{\omega_1}}\ket{H_{\omega_2}}).
\end{split}
\label{eq:clockwiseHV}
\end{equation}
The polarizing beam splitter then sorts the orthogonal polarization states into two distinct spatial modes $a$ and $b$:
\begin{equation}
\begin{split}
\ket{\psi}_{\text{CW}}\rightarrow\frac{1}{\sqrt{2}}(\ket{H_{\omega_1}}_b\ket{V_{\omega_2}}_a-\ket{V_{\omega_1}}_a\ket{H_{\omega_2}}_b).
\end{split}
\label{eq:clockwise}
\end{equation}
Analogously, for the counter-clockwise direction of the Sagnac loop (see Fig.~\hyperref[figure1]{\ref*{figure1}(b)}), one obtains bi-photons in state
\begin{equation}
\begin{split}
\ket{\psi}_{\text{CCW}}=\frac{1}{\sqrt{2}}(\ket{H_{\omega_1}}_a\ket{V_{\omega_2}}_b-\ket{V_{\omega_1}}_b\ket{H_{\omega_2}}_a).
\end{split}
\label{eq:counterclockwise}
\end{equation}
Superimposing the two states $\ket{\psi}_{\text{CCW}}$ and $\ket{\psi}_{\text{CW}}$ results in a maximally polarization-frequency hyperentangled state
\begin{equation}
\begin{split}
\ket{\Psi^+_\text{p}}\otimes\ket{\Psi^-_\omega}=\frac{1}{2}(\ket{HV}+e^{i\varphi_\text{p}}\ket{VH})\otimes(\ket{\omega_1\omega_2}-\ket{\omega_2\omega_1}),
\end{split}
\label{eq:hyperentanglement}
\end{equation}
where $\varphi_\text{p}$ is the phase of the pump polarization state, which is set outside of the Sagnac loop.

Since only pairs of photons anti-parallel in their polarization with respect to the rectilinear reference frame of the PBS ($\ket{HV}$ or $\ket{VH}$) are routed into two separate output ports, the unwanted polarization-parallel contributions ($\ket{HH}$ or $\ket{VV}$) can be eliminated by post-selecting on coincidences between two distinct spatial modes. Thus, the PBS in the Sagnac interferometer actually acts as state purification to improve the fidelity of the polarization-entangled state.

\section{Experimental characterization of the hyperentangled state}
\begin{figure}[!t]
\centering
\includegraphics[width=\linewidth]{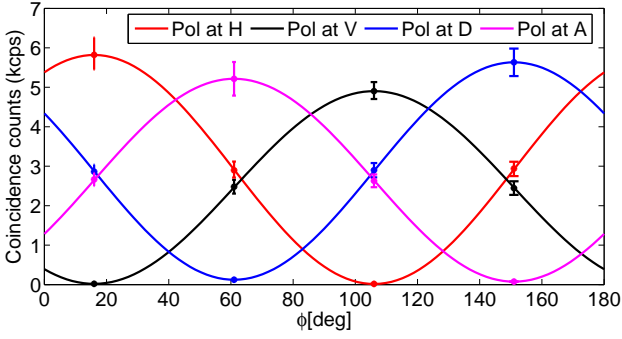}
\caption{Correlations in the polarization subspace. The two-fold coincidence counts are measured in two mutually unbiased D/A and H/V bases. All error bars in experimental data are estimated by statistical methods assuming a Poisson distribution.}
\label{figure_2}
\end{figure}
Akin to our previous experimental setup in Ref. \cite{chen2018polarization}, the hyperentanglement source is implemented by pumping a pair of crossed 10-mm-long ppKTP crystals with a grating stabilized laser emitting continuous wave at wavelength of $\unit[405]{nm}$. The pump beam is set to be linearly polarized at $45^\circ$ with respect to the reference frame of the PBS, making the SPDC process to occur with equal probability in clockwise and counter-clockwise directions. To achieve the desired diagonal and anti-diagonal polarizations, we design a V-groove oven such that two nonlinear crystals oriented along the oven are phase-matched with diagonally or anti-diagonally polarized photons, respectively. By superimposing down-converted photons emitted from both propagation directions on a PBS, they are sorted into distinct spatial modes deterministically (see Appendix A). Our source produces hyperentangled photon pairs at a rate of 4.4 kcps per mW of pump power with a symmetric heralding efficiency of $17\%$. Without any bandpass filtering this corresponds to a spectral brightness of 8.3 kcps/nm per mW of pump power.

We verify the quantum correlations of the produced state successively in the polarization and frequency subspace. For the polarization degree of freedom we certify entanglement in an assumption-free manner while for the frequency degree of freedom we present two methods with different levels of assumptions about the state. In the next section, we combine these results to prove the generation of high-dimensional entanglement.

In order to verify the entanglement in the polarization domain, we measure two-photon correlations in two mutually unbiased bases, yielding interference visibilities of $V_{H/V}=99.3\pm0.3\%$ in the H/V basis and $V_{A/D}=96.4\pm0.5\%$ in the A/D basis (see Fig.\ \ref{figure_2}). These visibilities imply lower bounds of $F_\text{p} \geq 0.979$ and $C_\text{p} \geq 0.958$ on the Bell-state fidelity and concurrence, respectively.

\begin{figure}[!t]
\centering
\subfigure[]{
\label{Fig3.sub.1}
\includegraphics[width=0.48\linewidth]{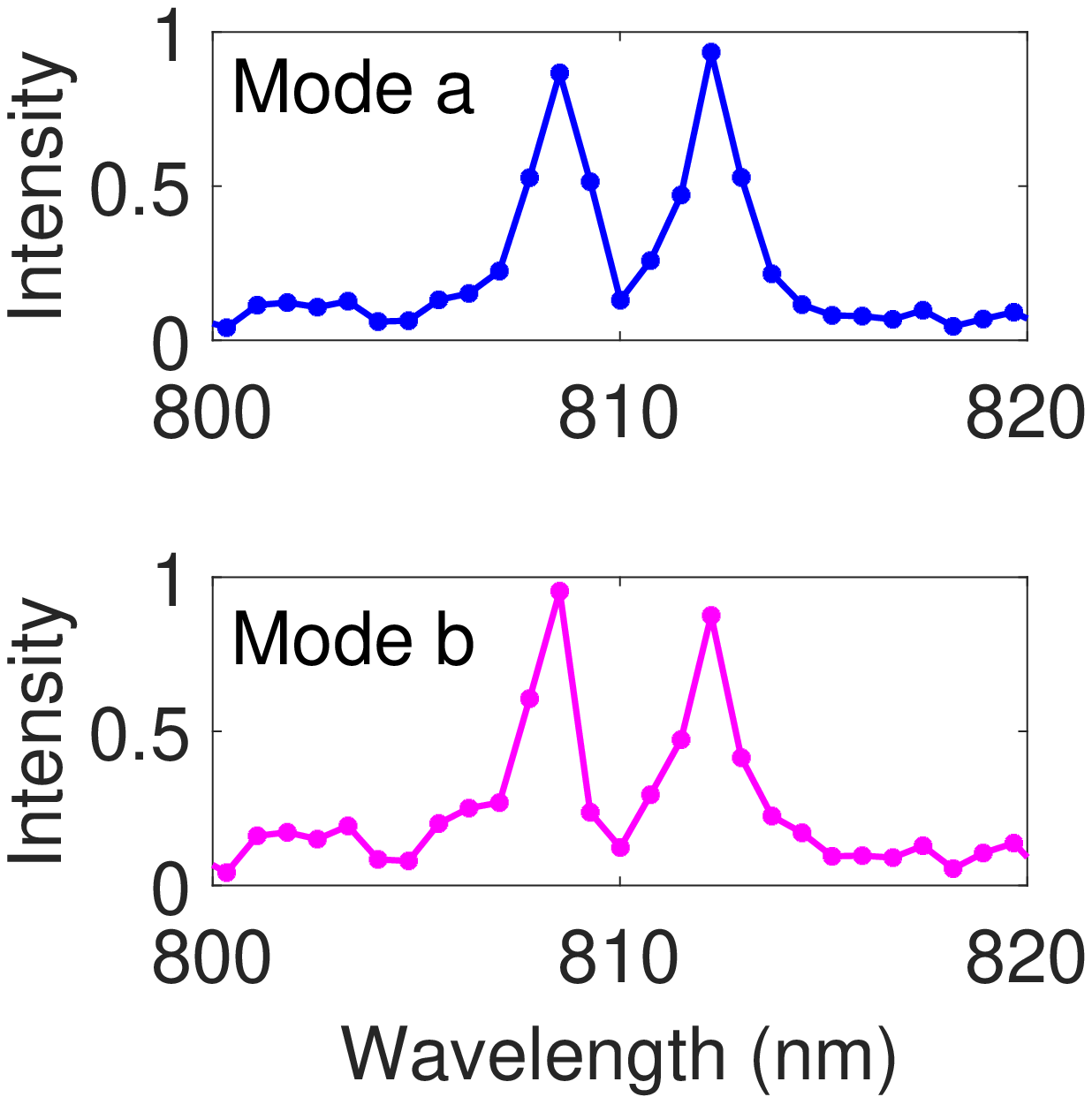}}
\subfigure[]{
\label{Fig3.sub.2}
\includegraphics[width=0.48\linewidth]{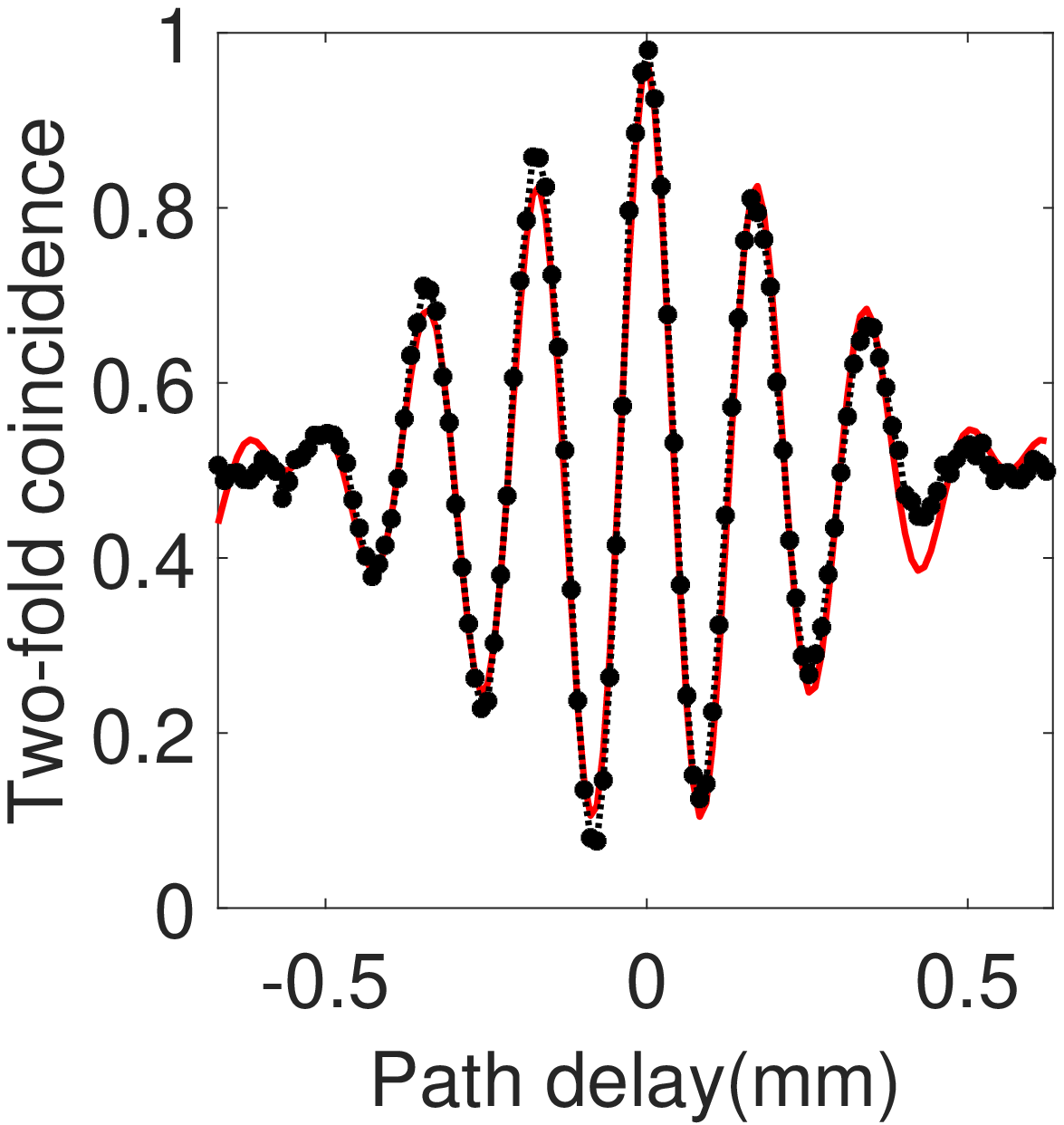}}
\subfigure[]{
\label{Fig3.sub.3}
\includegraphics[width=0.48\linewidth]{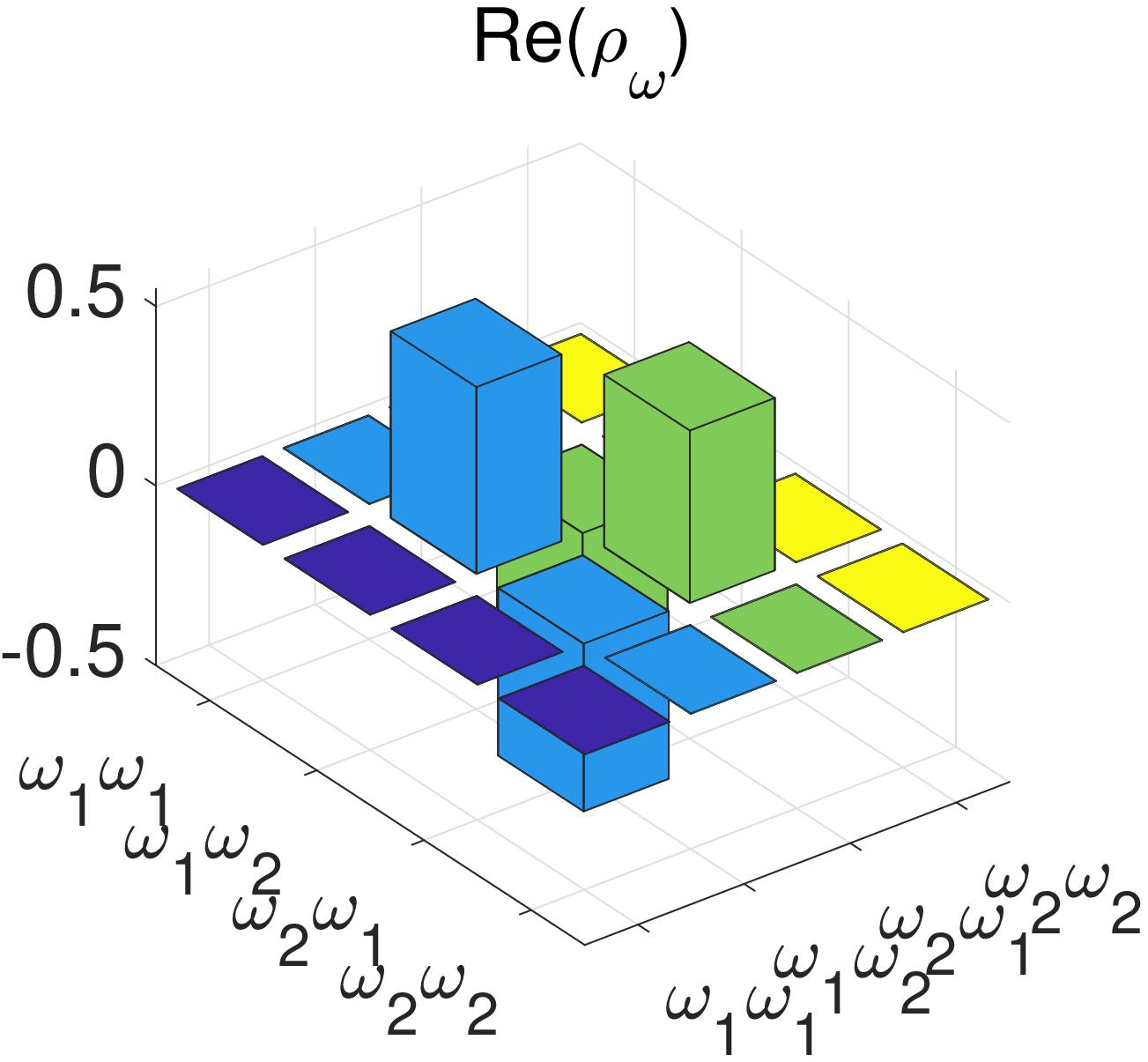}}
\subfigure[]{
\label{Fig3.sub.4}
\includegraphics[width=0.48\linewidth]{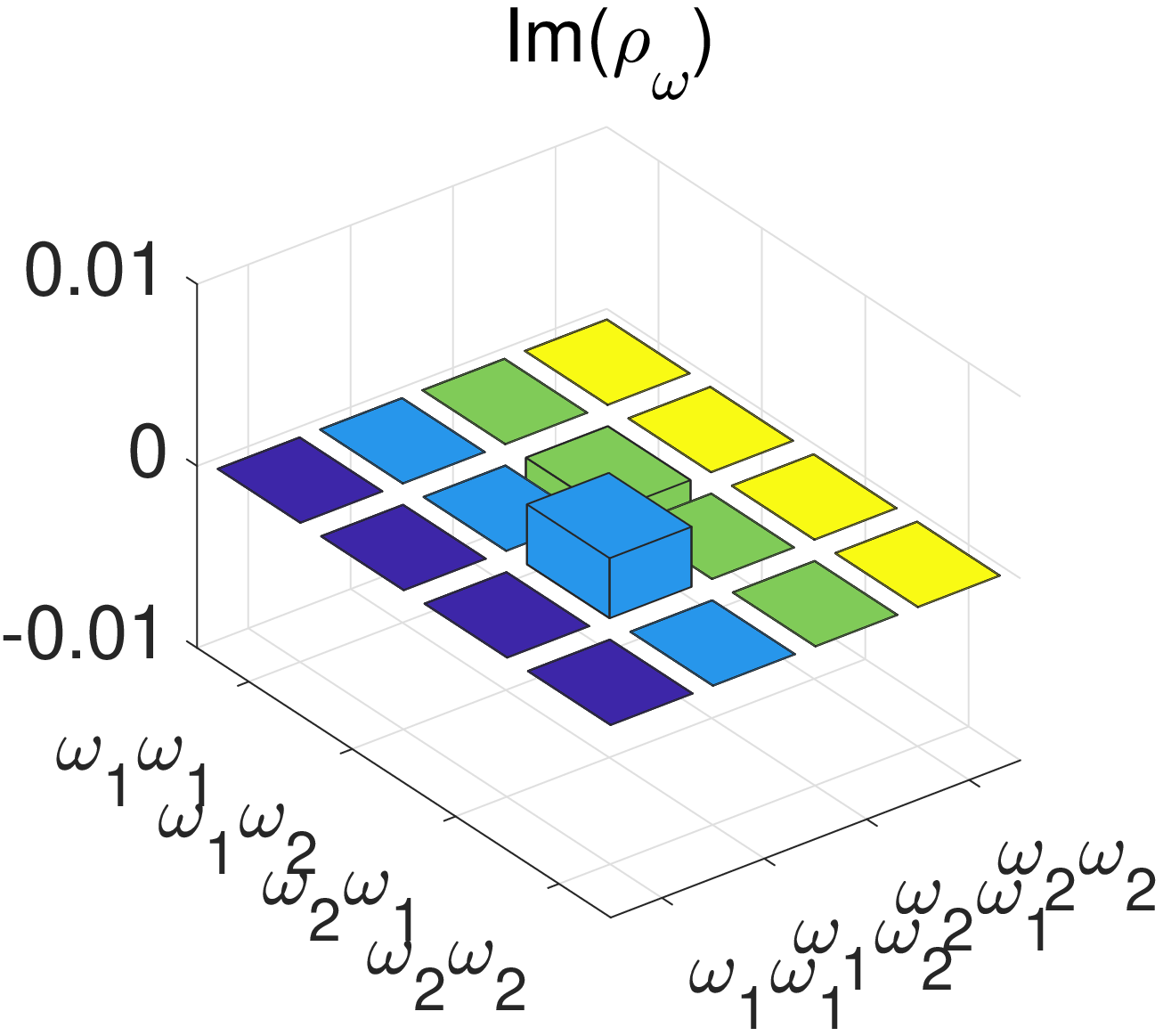}}
\caption{Correlations in the frequency subspace. (a) Spectral distribution of the two spatial modes observed by a single-photon spectrometer. (b) Normalized coincidence rate after the Hong-Ou-Mandel interferometer as a function of the relative path delay . (c) Real and (d) imaginary part of estimated restricted density matrix.}
\label{figure_3}
\end{figure}

The verification of entanglement in the discrete frequency subspace is more elaborate due to the difficulty of a mutually unbiased measurement in the frequency domain. In order to separate the polarization from the frequency domain, polarizers are placed before the frequency analysis as depicted in Fig.~\hyperref[figure1]{\ref*{figure1}(d)}. The existence of two separated frequency bins in each spatial mode is verified by a single-photon spectrometer (see Fig.\ \ref{Fig3.sub.1}), which shows a good overlap of the spectra in both spatial modes. While a non-local measurement of the coherence of frequency-entangled states is difficult without the assistance of nonlinear optical process \cite{tanzilli2005photonic} and a time-resolved measurement \cite{guo2017testing}, it can be quantified utilizing spatial beating in HOM interference \cite{ou2007multi,jin2016simple,xie2015harnessing}. This non-classical beating can be observed by scanning the time-of-arrival of one of the photons incident on the 50:50 beam splitter, which constitutes a HOM interferometer. The corresponding interference fringes can be observed in the two-fold coincidences between the two output ports of the beam splitter (see Fig. \ref{Fig3.sub.2}). As a consequence of the anti-symmetry of the state $\ket{\Psi^-_\omega}$, we can observe photon-antibunching at zero path delay. Discrete frequency-entanglement manifests itself in sinusoidal oscillations of the interference fringes within a Gaussian envelope as a function of relative time delay $\tau$. This can be modelled with a coincidence probability of \cite{ramelow2009discrete,fedrizzi2009anti}
\begin{equation}
\begin{split}
p_c(\tau)=\frac{1}{2}-\frac{V_\omega}{2}cos(\mu\tau+\varphi_\omega)(1-|\frac{2\tau}{\tau_c}|) \quad \text{for} |\tau|<\frac{\tau_c}{2},
\end{split}
\label{eq:frequency probability}
\end{equation}
where $\tau$ is relative arrival time delay of two photons at the beam splitter, $\tau_c$ is the single photon coherence time that equals the base-to-base envelope width, and $\mu=\omega_2-\omega_1$ is the detuning of two well separated frequency bins. The magnitude of the oscillations is parametrized with the visibility $V_\omega$, while $\varphi_\omega$ is a phase-offset. A fit of our measurement data to \eqref{eq:frequency probability} reveals the parameters of the restricted density matrix $\rho_\omega$, which reads
\begin{equation}
\rho_\omega=\left\{
\begin{matrix}
0&0&0&0\\
0&p_\omega&\frac{V_\omega}{2}e^{-i\varphi_\omega}&0\\
0&\frac{V_\omega}{2}e^{i\varphi_\omega}&1-p_\omega&0\\
0&0&0&0\\
\end{matrix}
\right\}
\label{eq:frequency matrix}
\end{equation}
in the computational basis $\{\ket{\omega_1\omega_1}$,  $\ket{\omega_1\omega_2}$, $\ket{\omega_2\omega_1}$,  $\ket{\omega_2\omega_2}\}$ \cite{fickler2012quantum}. Outside of the inner 2x2 submatrix, the density matrix elements are set to zero, because of energy conservation in the process of SPDC with a narrow-band pump laser. The balance parameter $p_\omega$ and the visibility $V_\omega$ satisfy the physical constraints $0\leq p_\omega\leq1$ and $0\leq\frac{V_\omega}{2}\leq\sqrt{p_\omega(1-p_\omega)}$.

Based on our measurement results, we estimate a coherence time of $\tau_c\approx\unit[3.8]{ps}$, which is inversely related to a single-photon frequency bandwidth of $\Delta f_{\text{FWHM}}\approx\unit[0.24]{Thz}$ or a wavelength bandwidth of $\Delta \lambda_{\text{FWHM}}\approx\unit[0.53]{nm}$. The frequency detuning $\mu\approx\unit[1.75]{THz}$ is much larger than $\Delta f_{\text{FWHM}}$, which again confirms the separation of the two frequency bins. The resulting visibility is $V_\omega\approx94.3\%$, while the relative phase is estimated to be $\varphi_\omega\approx \unit[179.6]{^\circ}$, which is close to $\pi$. The balance parameter is calculated from the single-photon spectra of Fig.\ \ref{Fig3.sub.1}, resulting in $p_\omega\approx0.52$. Thus we are able to estimate the restricted density matrix $\rho_\omega$ as depicted in Fig.\ \ref{Fig3.sub.3} and \ref{Fig3.sub.4}. The fidelity to the Bell-state $\ket{\Psi^-_\omega}$ follows from $F_\omega=\mathbf{Tr}(\rho_\omega\ket{\Psi^-_\omega}\bra{\Psi^-_\omega})\approx0.971$, which implies a frequency subspace concurrence of $C_\omega\approx0.942$.

In order to demonstrate the versatility of our source in the frequency domain, we changed the detuning of the frequeny bins $\mu$ by increasing the temperature of the nonlinear crystal, while monitoring the fidelity to the polarization Bell state $\ket{\Psi^+_\text{p}}$.
For instance, we observed a fidelity of $\sim 0.965$ in the scenario of $\mu=\unit[7.35]{THz}$ by setting temperature at $\unit[50]{^\circ C}$, and a fidelity of $\sim 0.958$ in the scenario of $\mu=\unit[14.12]{THz}$ by setting temperature at $\unit[85]{^\circ C}$. Moreover, the measured photon pair rates are almost constant irrespective of the frequency detuning.

The preceding verification of entanglement in the frequency subspace is only valid under the assumption that the visibility $V_\omega$, which is extracted from a fit of the experimental data to Eq.~\eqref{eq:frequency probability}, indeed corresponds to $V_\omega$ in the restricted density matrix, Eq.~\eqref{eq:frequency matrix}.
In order to provide a stronger form of entanglement verification, we now derive a lower bound for the fidelity in the discrete frequency subspace that relies only on the assumption of energy conservation but does not require any other constraints to be imposed on the density matrix.

This lower bound is derived as a function of the experimentally measured visibilty, which as shown in Appendix \ref{a:boundderivation}, is given by
\begin{equation}
V_\omega= \frac{2|\bra{\omega_1\omega_2}\rho\ket{\omega_2\omega_1}|}{\bra{\omega_1\omega_2}\rho\ket{\omega_1\omega_2} +\bra{\omega_2\omega_1}\rho\ket{\omega_2\omega_1}}.
\end{equation}

Now, the fidelity $F_\omega$ of $\rho$ with the maximally entangled state $\ket{\Psi^-_\omega}$, after optimizing over the path delay, is shown in Appendix \ref{a:boundderivation} to be
\begin{align}
F_\omega =& \frac{1}{2} \Big(\bra{\omega_1\omega_2}\rho\ket{\omega_1\omega_2} +\bra{\omega_2\omega_1}\rho\ket{\omega_2\omega_1}\Big)+ |\bra{\omega_1\omega_2}\rho\ket{\omega_2\omega_1}| \\
\geq& \ 2|\bra{\omega_1\omega_2}\rho\ket{\omega_2\omega_1}|.
\end{align}

Assuming energy conservation, which implies $\bra{\omega_1\omega_2}\rho\ket{\omega_1\omega_2} +\bra{\omega_2\omega_1}\rho\ket{\omega_2\omega_1} = 1$, we arrive at the lower bound of
\begin{equation}
F_\omega \geq V_\omega.
\end{equation}
All details are outlined in Appendix \ref{a:boundderivation}.

This results in a measured lower bound of the fidelity in the discrete frequency subspace of $F_\omega\geq0.855$ that relies exclusively on the assumption of energy conservation, a stronger result than our previous method.

The measured high fidelities of the reduced two-qubit states in both the polarization and frequency subspaces, with respect to a maximally entangled two-qubit state, indicate the presence of bipartite entanglement in both subspaces and its absence on the global state across the frequency-polarization partition -- suggesting the presence of hyper-entanglement. In the following, we show that this is indeed the case by verifying the generation of high-dimensional entanglement.

\section{Verification of high-dimensional entanglement}
Having estimated the value of both the polarization and frequency subspace fidelities, we can now infer entanglement properties of the two-ququart global state enconded jointly in the polarization and frequency DOF. In order to do so, we formulate an optimization problem, in the same fashion as Ref. \cite{steinlechner2017distribution}. Namely, we search for a global state of two ququarts whose reduced two-qubit states satisfy the properties we have experimentally measured, i.e., have values for the fidelity with respect to a maximally entangled two-qubit state that are equal to the ones that were measured. Among all possible two-ququart states that have subspace fidelities compatible with the measured ones, we must choose the one with lowest fidelity with respect to a maximally entangled two-ququart state, in order not to overestimate the entanglement of the global state. Consequently, the fidelity of the optimization state will constitute a lower bound for the fidelity of the experimental state.

This problem can be efficiently solved via semidefinite programming (SDP), a class of convex optimisation problems. Let $\rho_{\text{p}_A\omega_A\text{p}_B\omega_B}$ be the global $4$x$4$-dimensional state composed of two polarization qubits and two frequency qubits, shared by parties A and B, which are the recipients of photons in spatial mode $a$ and $b$, respectively. Let $F_\text{p}$ be the fidelity of the reduced polarization state, $\rho_{\text{p}_A\text{p}_B}=\text{Tr}_{\omega_A\omega_B}(\rho_{\text{p}_A\omega_A\text{p}_B\omega_B})$, with respect to a $2$x$2$-dimensional maximally entangled state and $F_\omega$ be the fidelity of the reduced frequency state, $\rho_{\omega_A\omega_B}=\text{Tr}_{\text{p}_A\text{p}_B}(\rho_{\text{p}_A\omega_A\text{p}_B\omega_B})$, also with respect to a $2$x$2$-dimensional maximally entangled state. Then, a lower bound for the fidelity $F_{\text{p}\omega}$ of the global state $\rho_{\text{p}_A\omega_A\text{p}_B\omega_B}$ with respect to a $4$x$4$-dimensional maximally entangled state is given by:
\begin{align}
\begin{split}
\text{given}            &\hspace{0.2cm}F_\text{p}, F_\omega\\
F_{p\omega} \geq \min	&\hspace{0.2cm} \text{Tr}(\rho_{\text{p}_A\omega_A\text{p}_B\omega_B}\ketbra{\Phi^{+}_{4}}{\Phi^{+}_{4}})\\
\text{s.t.}				&\hspace{0.2cm} F_\text{p} = \text{Tr}(\rho_{\text{p}_A\text{p}_B}\ketbra{\Phi^{+}_{2}}{\Phi^{+}_{2}}),\\
		 				&\hspace{0.2cm} F_\omega = \text{Tr}(\rho_{\omega_A\omega_B}\ketbra{\Phi^{+}_{2}}{\Phi^{+}_{2}}),\\
						&\hspace{0.2cm} \rho_{\text{p}_A\omega_A\text{p}_B\omega_B} \geq 0, \ \text{Tr}(\rho_{\text{p}_A\omega_A\text{p}_B\omega_B})=1,
\end{split}
\end{align}
where $\ket{\Phi^{+}_{d}}=\frac{1}{\sqrt{d}}\sum_{i=1}^d\ket{ii}$ \cite{steinlechner2017distribution}.

We solve this problem for a polarization subspace fidelity of $F_\text{p} = 0.979$, once using the frequency subspace fidelity obtained with our model of $F_\omega = 0.971$, and once for the frequency subspace fidelity obtained only assuming energy conservation of $F_\omega = 0.855$. The solution yields the lower bounds of $F_{\text{p}\omega}\geq 0.950$ and $F_{\text{p}\omega}\geq 0.834$, for each method respectively.

From the fidelity of the global state it is possible to estimate the dimensionality of its entanglement. Entanglement dimensionality is a quantifier that represents the minimum number of levels one needs to faithfully represent the state and its correlations in any global product basis. A lower bound for the entanglement dimensionality $d_\text{ent}$ of a $d$x$d$-dimensional state that has a fidelity $F$ with respect to the maximally entangled state is given by

\begin{equation}
d_\text{ent} \geq \left \lceil{d F}\right \rceil,
\end{equation}
where $\left \lceil{}\right \rceil$ is the ceiling function. We refer to Ref.~\cite{Fickler2014} or \cite{Friis2019review} for a detailed proof.
Using the above relation, from the fidelity lower bound of $F_{\text{p}\omega}\geq 0.950$ obtained from the first method we certify $d_\text{ent}=4$. Similarly, from the fidelity lower bound of $F_{\text{p}\omega}\geq 0.834$ we also certify $d_\text{ent}=4$, an even stronger result since it is achieved with fewer assumptions on the preparation of the state. This concludes the proof that high-dimensional entanglement has indeed been produced in our setup via hyper-entanglement.

\section{Discussion}
Quantum interference is a versatile tool in the quantum engineering toolbox. Here we make dual use of this phenomenon; both to \emph{generate} a polarization discrete-frequency hyperentangled state  without the usual requirement for detection post-selection and to \emph{analyse} high-dimensional entanglement stored in two independent degrees of freedom. The verification of high-dimensional entanglement further verifies the quality of the produced hyperentangled state.

Entanglement in multiple degrees of freedom enables us to encode many qubits into different properties of single photons. Since the hyper-entangled state prepared by our approach can be readily transformed into polarization-frequency cluster states, we hope that our work can pave the avenue for one-way quantum computation. Here, the challenge of implementing multi-qubit gates is shifted to the capability of creating cluster states \cite{ciampini2016path}. Furthermore, independent measurements in polarization and frequency DoFs may be sufficient for entanglement witness of hyper-entangled cluster states by using analogous methodology \cite{toth2005entanglement}. Its quality is therefore well verified, making it an ideal candidate for more complex quantum applications.

The versatility of our approach enables its extension to other platforms, such as optical waveguides or integrated photonics. We hope that our work inspires experiments which harness quantum interference to engineer hyperentangled states in other photonic degrees of freedom, such as orbital angular momentum, thereby setting the stage for quantum information processing in evermore complex quantum systems.

\section*{Acknowledgements}
Financial support from the Austrian Research Promotion Agency (FFG) Projects - Agentur f\"{u}r Luft- und Raumfahrt (FFG-ALR contract No. 6238191 and No. 866025), the European Space Agency (ESA contract No. 4000112591/14/NL/US), the Austrian Science Fund (FWF) through the START project Y879-N27, as well as the Austrian Academy of Sciences is gratefully acknowledged.
\bibliography{apssamp}

\appendix
\section{Experimental setup}
\begin{figure}[h]
\centering
\includegraphics[width=\linewidth]{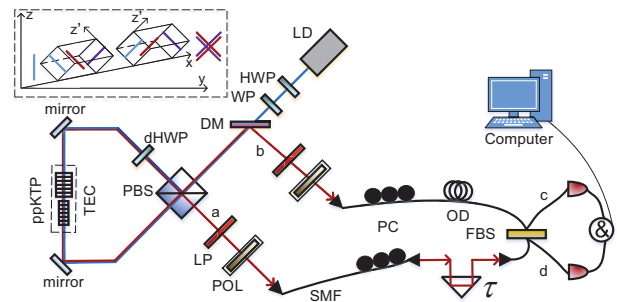}
\caption{Experimental setup of polarization-frequency hyperentanglement source. LD: laser diode; PBS: polarizing beam splitter; HWP: half wave plate; WP: wave plate; DM: dichroic mirror; ppKTP: type-II periodically poled potassium titanyl phosphate crystal; TEC: temperature controller; LP: long pass filter, POL: polarizer. The top-left inset illustrates that the design of V-groove oven with inclination of $45^\circ$ along the optical axis enables the generation of photon pairs with diagonal or anti-diagonal polarization. The PBS and HWP inside the Sagnac loop are operated at dual-wavelength of $\unit[405/810]{nm}$.}
\label{fig_S1}
\end{figure}
The experimental setup of our polarization and discrete frequency hyperentanglement source is depicted in Fig.\ \ref{fig_S1}. We generate the two-photon states in the form of Eq. (\textcolor{blue}{1}) by pumping a pair of crossed ppKTP crystals with a grating-stabilized laser diode emitting continuous wave at a wavelength of $\unit[405]{nm}$ (Toptica DL Pro). Through a PBS and a zero-order HWP with its optical axis oriented at $22.5^\circ$, the pump beam is set to be linearly polarized at $45^\circ$ with respect to the reference frame of the PBS, making the spontaneous parametric down conversion (SPDC) process to occur with equal probability in clockwise and counter-clockwise directions. To achieve the desired diagonal and anti-diagonal polarizations, we designed a V-groove oven such that two crossed crystals are oriented along the oven as shown in the inset of Fig.\ \ref{fig_S1}. The nonlinear crystals are placed flat inside the oven, which means they are phase-matched for SPDC with diagonally or anti-diagonally polarized photons, respectively. The crossed crystals scheme utilizes two mutually orthogonally oriented 10-mm-long ppKTP crystals. They are manufactured for type-II collinear phase matching with pump (p), signal (s) and idler (i) photons at approximately center wavelengths of $\lambda_{p}\approx\unit[405]{nm}$ and $\lambda_{s,i}\approx\unit[810]{nm}$ at a crystal temperature of $33^\circ C$. A dual-wavelength HWP is added to compensate the phase difference caused by different group velocities of pump beam and down-converted photons in ppKTP crystals. By superimposing down-converted photons emitted from both propagation directions on a PBS, they are sorted into distinct spatial modes deterministically. Then, the down converted signal and idler photons are separated from the pump beam by using a dichroic mirror. Two long-pass filters are used to eliminate the remaining pump and background photons. In order to erase spatial $``which-crystal''$ information, the down converted photons are coupled into single mode fiber.

For analyzing the polarization correlation of the hyperentangled state, we measure two-photon interference in two mutually unbiased bases assisted by polarizers prior to detection. For analyzing the frequency correlation of hyperentanglement, we observe the spatial beating of nonoverlapping optical frequencies by scanning the arriving time of two photons at a balanced beam splitter, which constitutes a Hong-Ou-Mandel interferometer. Then the down converted photons are detected by silicon avalanche photondiodes, and two-fold coincidence events are identified with a time window of $\sim \unit[3]{ns}$. The measurement results enable us to characterize polarization and frequency entanglement in independent subspaces.

\section{HOM interference for frequency entanglement}
\begin{figure}[h]
\centering
\includegraphics[width=0.6\linewidth]{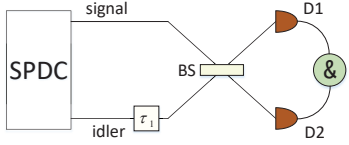}
\caption{HOM interference of frequency entanglement}
\label{fig_S2}
\end{figure}
Here, we simply demonstrate the process for HOM interference of two dimensional frequency entanglement. The basic schematic is depicted in Fig.\ \ref{fig_S2}. The two-photon state from a SPDC process can be described as
\begin{equation}
\ket{\psi}=\int_0^\infty\int_0^\infty d\omega_sd\omega_if(\omega_s,\omega_i)\hat{a}_s^\dag(\omega_s)\hat{a}_i^\dag(\omega_i)\ket{0}.
\end{equation}
The operation of a balanced beam splitter can be expressed as
\begin{equation}
\begin{split}
\hat{a}_s^\dag(\omega_s)=\frac{1}{\sqrt{2}}[\hat{a}_1^\dag(\omega_s)+\hat{a}_2^\dag(\omega_s)]\\
\hat{a}_i^\dag(\omega_s)=\frac{1}{\sqrt{2}}[\hat{a}_1^\dag(\omega_i)-\hat{a}_2^\dag(\omega_i)].\\
\end{split}
\end{equation}
As we introduce a tunable time delay $\tau_1$, it generates phase shift $exp(-i\omega_i\tau_1)$ to the idler photon with respect to the signal photon. Thus, after the operation of 50:50 beam splitter, we get two-photon state as
\begin{equation}
\begin{split}
\ket{\psi}=&\frac{1}{2}\int_0^\infty\int_0^\infty d\omega_sd\omega_if(\omega_s,\omega_i)e^{-i\omega_i\tau_1}[i\hat{a}_1^\dag(\omega_s)\hat{a}_1^\dag(\omega_i)\\
&+i\hat{a}_2^\dag(\omega_s)\hat{a}_2^\dag(\omega_i)+\hat{a}_1^\dag(\omega_i)\hat{a}_2^\dag(\omega_s)-\hat{a}_1^\dag(\omega_s)\hat{a}_2^\dag(\omega_i)]\ket{0},
\end{split}
\end{equation}
where subscript 1/2 represent two output modes of the beam splitter. For the post-selected coincidence counts by two detectors, only the last two terms of Eq. (\textcolor{blue}{B3}) are non-vanishing. So it can be simplified to
\begin{equation}
\begin{split}
\ket{\psi}=&\frac{1}{2}\int_0^\infty\int_0^\infty d\omega_sd\omega_if(\omega_s,\omega_i)e^{-i\omega_i\tau_1}\\
&[\hat{a}_1^\dag(\omega_i)\hat{a}_2^\dag(\omega_s)-\hat{a}_1^\dag(\omega_s)\hat{a}_2^\dag(\omega_i)]\ket{0}\\
=&\frac{1}{2}\int_0^\infty\int_0^\infty d\omega_sd\omega_i[f(\omega_s,\omega_i)e^{-i\omega_i\tau_1}\\
&-f(\omega_i,\omega_s)e^{-i\omega_s\tau_1}]\hat{a}_1^\dag(\omega_s)\hat{a}_2^\dag(\omega_i)\ket{0}.\\
\end{split}
\end{equation}
As two photons after beam splitter are indistinguishable, we substitute $\omega_s$ and $\omega_i$ with $\omega_1$ and $\omega_2$. By multiplying $e^{-i\omega_i\tau_1}$ to cancel the global phase, we obtain
\begin{equation}
\begin{split}
\ket{\psi}=&\frac{1}{2}\int_0^\infty\int_0^\infty d\omega_1d\omega_2[f(\omega_1,\omega_2)-f(\omega_2,\omega_1)\\
&e^{-i(\omega_1-\omega_2)\tau_1}]\hat{a}_1^\dag(\omega_1)\hat{a}_2^\dag(\omega_2)\ket{0}.
\end{split}
\end{equation}
The detection operators of two detectors in different output modes are
\begin{equation}
\begin{split}
\hat{E}_1^{(+)}=\frac{1}{\sqrt{2\pi}}\int_0^\infty d\omega_1\hat{a}_1(\omega_1)e^{-i\omega_1t_1},\\
\hat{E}_2^{(+)}=\frac{1}{\sqrt{2\pi}}\int_0^\infty d\omega_2\hat{a}_2(\omega_2)e^{-i\omega_2t_2}.\\
\end{split}
\end{equation}
Thus we can calculate $\hat{E}_2^{(+)}\hat{E}_1^{(+)}\ket{\psi}$ as
\begin{equation}
\begin{split}
\hat{E}_2^{(+)}\hat{E}_1^{(+)}\ket{\psi}=&\frac{1}{2\pi}\int_0^\infty\int_0^\infty d\omega_1d\omega_2\hat{a}_1(\omega_1)\hat{a}_2(\omega_2)e^{-i\omega_1t_1}\\
&e^{-i\omega_2t_2}\times\frac{1}{2}\int_0^\infty\int_0^\infty d\omega_1^\prime d\omega_2^\prime[f(\omega_1^\prime,\omega_2^\prime)\\
&-f(\omega_2^\prime,\omega_1^\prime)e^{-i(\omega_1^\prime-\omega_2^\prime)\tau_1}]\hat{a}_1^\dag(\omega_1^\prime)\hat{a}_2^\dag(\omega_2^\prime)\ket{0}\\
=&\frac{1}{4\pi}\int_0^\infty\int_0^\infty d\omega_1d\omega_2[f(\omega_1,\omega_2)\\
&-f(\omega_2,\omega_1)e^{-i(\omega_1-\omega_2)\tau_1}]e^{-i\omega_1t_1}e^{-i\omega_2t_2}\ket{0},
\end{split}
\end{equation}
where we add $\omega^\prime$ to distinguish between the symbols for photon and detection frequency, albeit $\omega^\prime=\omega$. Finally the coincidence probability $P(\tau_1)$ as a function of time delay can be expressed as
\begin{equation}
\begin{split}
P(\tau_1)=&\langle\psi|\hat{E}_1^{(-)}\hat{E}_2^{(-)}\hat{E}_2^{(+)}\hat{E}_1^{(+)}\ket{\psi}\\
=&(\frac{1}{4\pi})^2\int_0^\infty\int_0^\infty d\omega_1d\omega_2[f(\omega_1,\omega_2)\\
&-f(\omega_2,\omega_1)e^{-i(\omega_1-\omega_2)\tau_1}]e^{-i\omega_1t_1}e^{-i\omega_2t_2}\\
&\times\int_0^\infty\int_0^\infty d\omega_1^\prime d\omega_2^\prime[f(\omega_1^\prime,\omega_2^\prime)\\
&-f(\omega_2^\prime,\omega_1^\prime)e^{-i(\omega_1^\prime-\omega_2^\prime)\tau_1}]^{\ast}e^{i\omega_1t_1}e^{i\omega_2t_2}\ket{0}\\
=&\frac{1}{4}\int_0^\infty\int_0^\infty d\omega_1d\omega_2|f(\omega_1,\omega_2)|^2+|f(\omega_2,\omega_1)|^2\\
&-2f(\omega_1,\omega_2)f(\omega_2,\omega_1)cos(\omega_1-\omega_2)\tau_1.
\end{split}
\end{equation}
If $f(\omega_1,\omega_2)$ is an even function, we have $f(\omega_1,\omega_2)=f(\omega_2,\omega_1)$, such that $P(\tau_1)$ exhibits a dip at position of $\tau_1=0$. On the other hand, if $f(\omega_1,\omega_2)$ is an odd function, we have $f(\omega_1,\omega_2)=-f(\omega_2,\omega_1)$ such that $P(\tau_1)$ exhibits a peak at the position of $\tau_1=0$.
\section{Verification of hyperentanglement after Hong-Ou-Mandel interference}
\begin{figure*}[!ht]
\centering
\subfigure[]{
\label{FigS3.sub.1}
\includegraphics[width=0.18\linewidth]{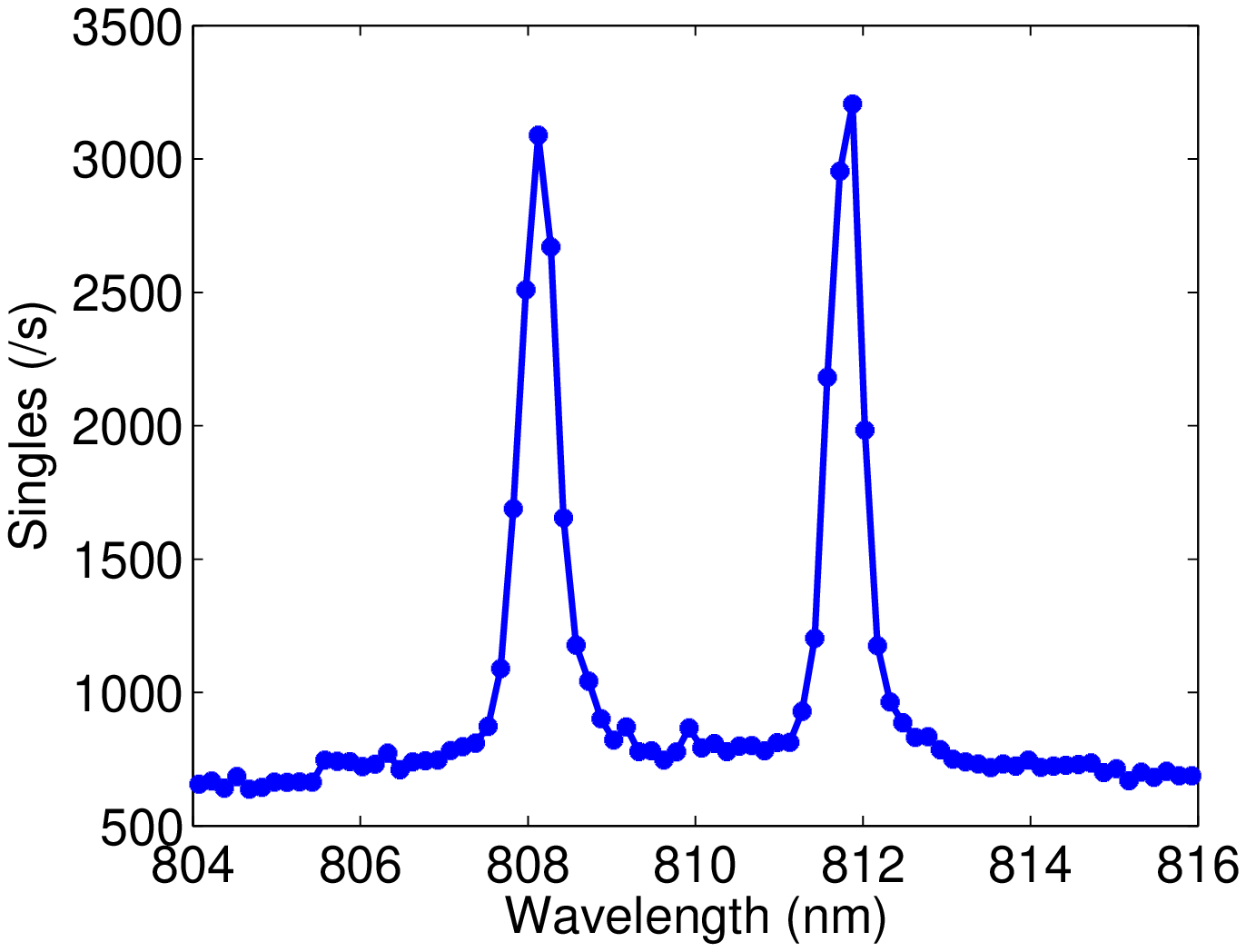}}
\subfigure[]{
\label{FigS3.sub.2}
\includegraphics[width=0.18\linewidth]{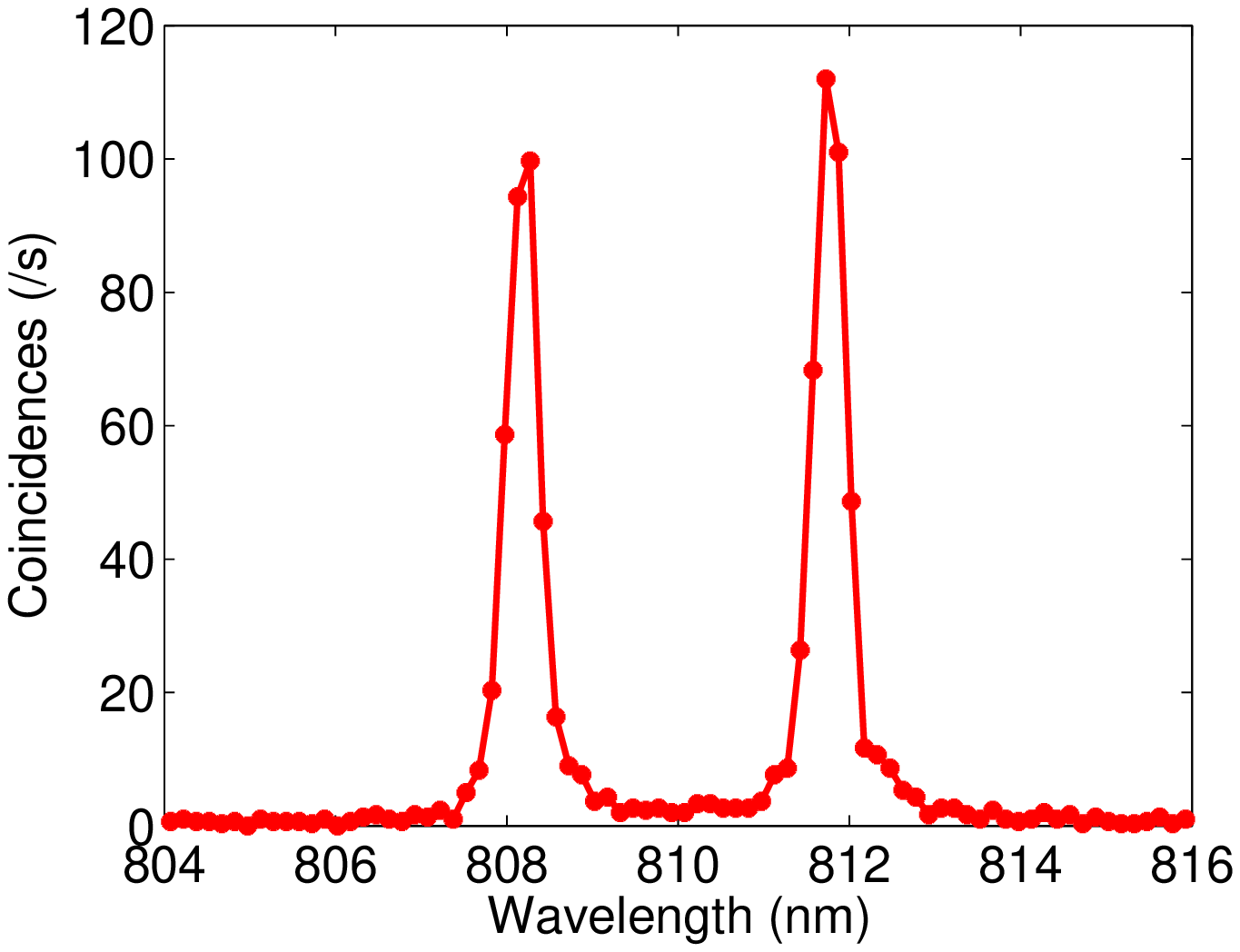}}
\subfigure[]{
\label{FigS3.sub.3}
\includegraphics[width=0.18\linewidth]{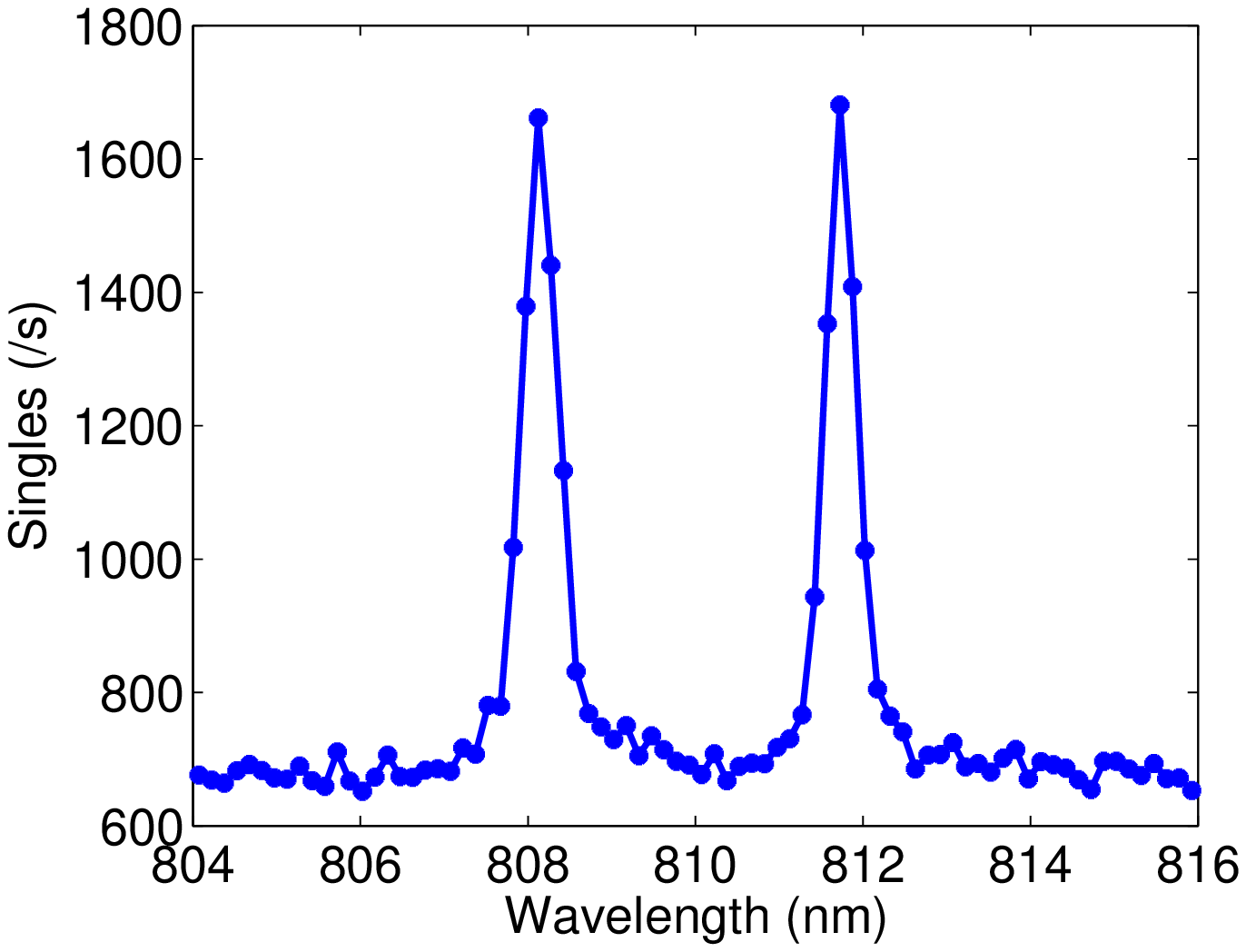}}
\subfigure[]{
\label{FigS3.sub.4}
\includegraphics[width=0.18\linewidth]{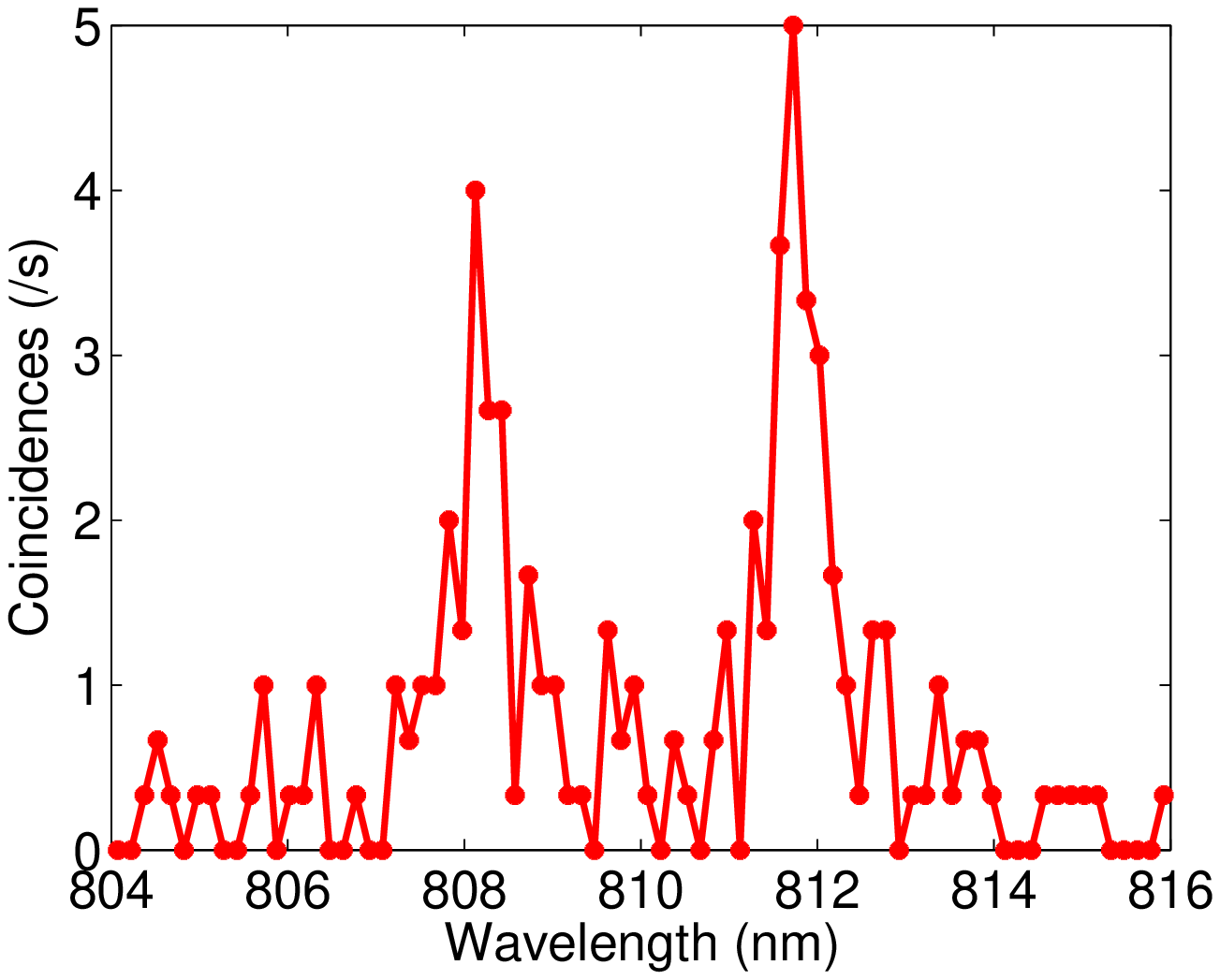}}
\subfigure[]{
\label{FigS3.sub.5}
\includegraphics[width=0.18\linewidth]{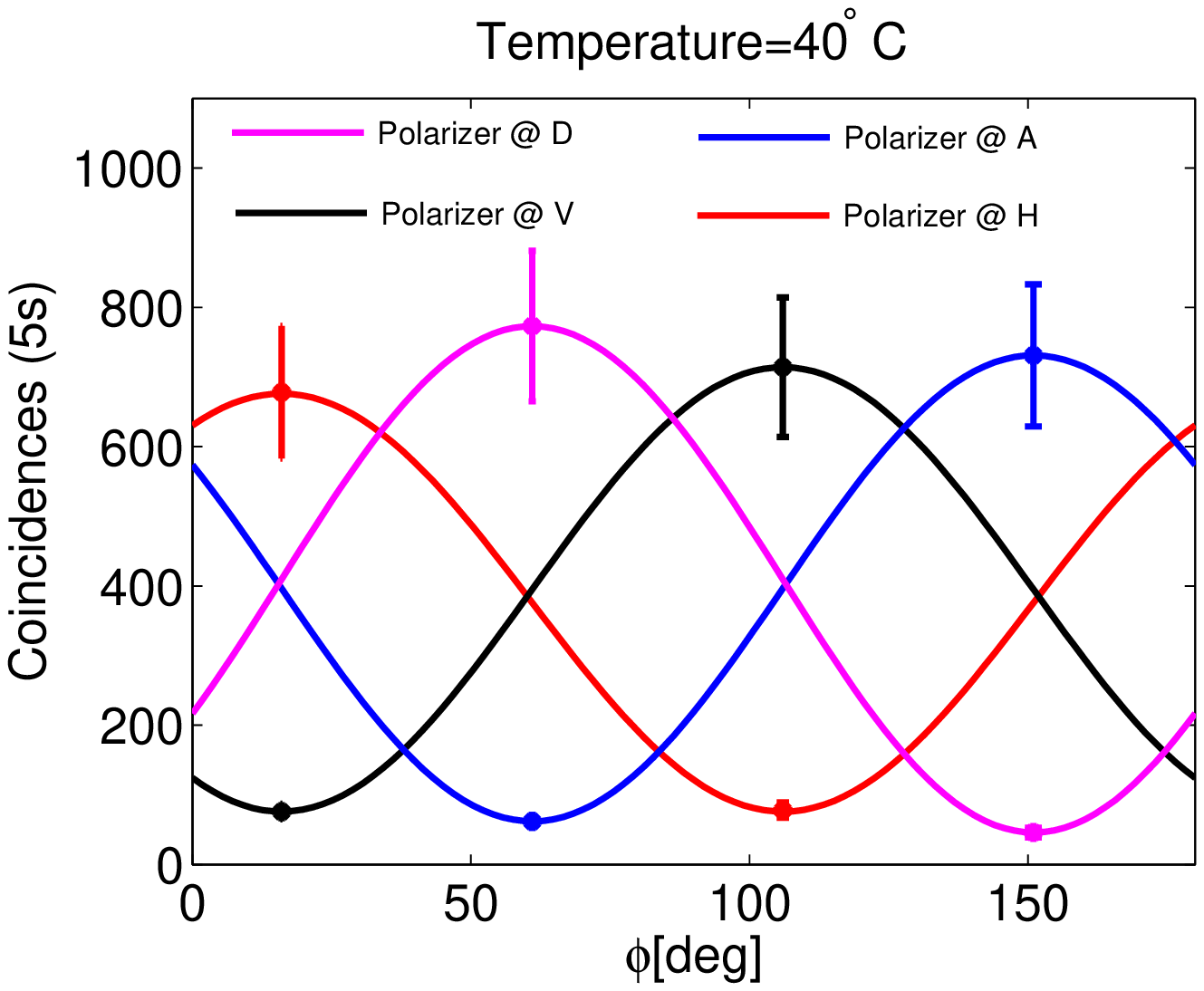}}
\subfigure[]{
\label{FigS3.sub.6}
\includegraphics[width=0.18\linewidth]{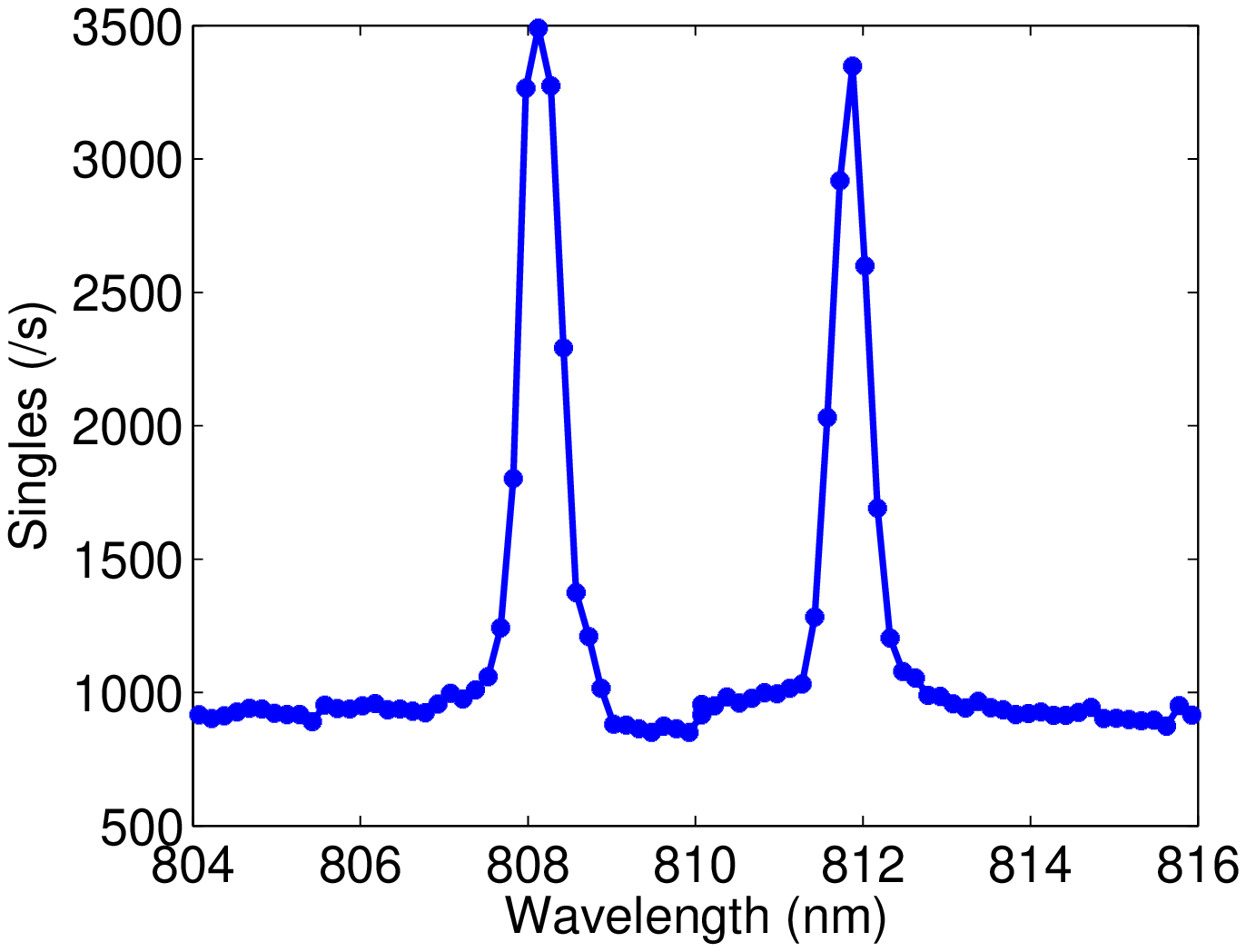}}
\subfigure[]{
\label{FigS3.sub.7}
\includegraphics[width=0.18\linewidth]{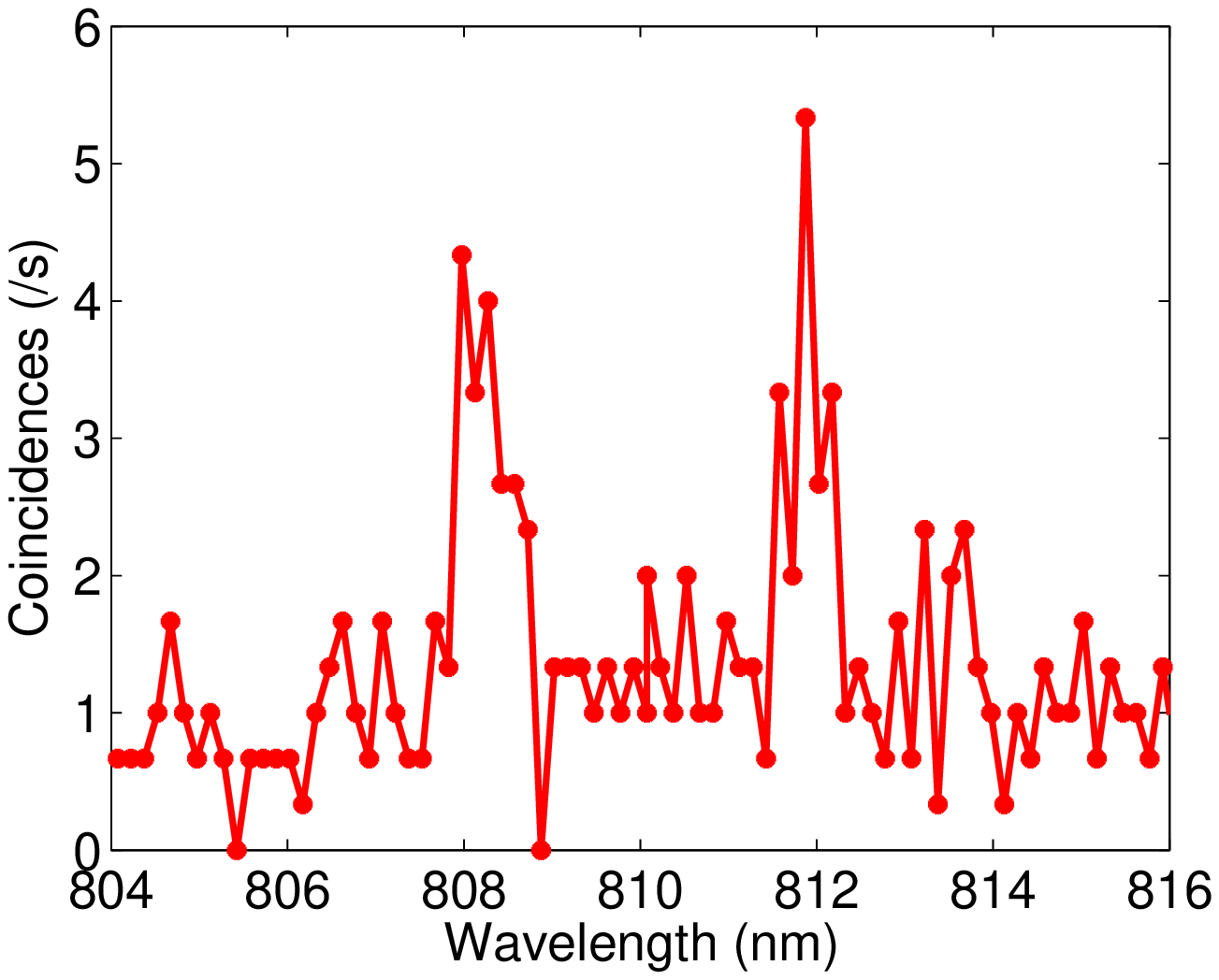}}
\subfigure[]{
\label{FigS3.sub.8}
\includegraphics[width=0.18\linewidth]{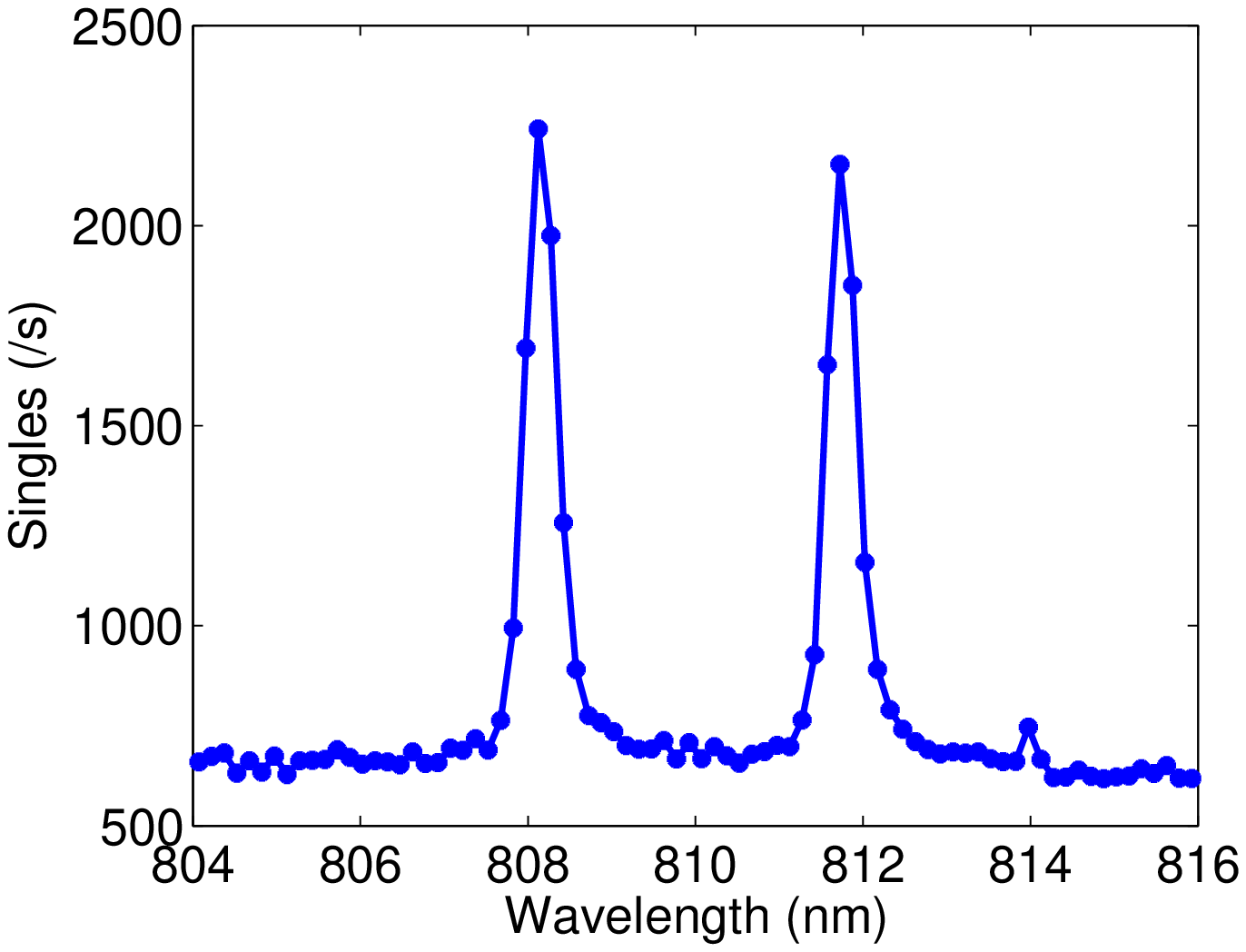}}
\subfigure[]{
\label{FigS3.sub.9}
\includegraphics[width=0.18\linewidth]{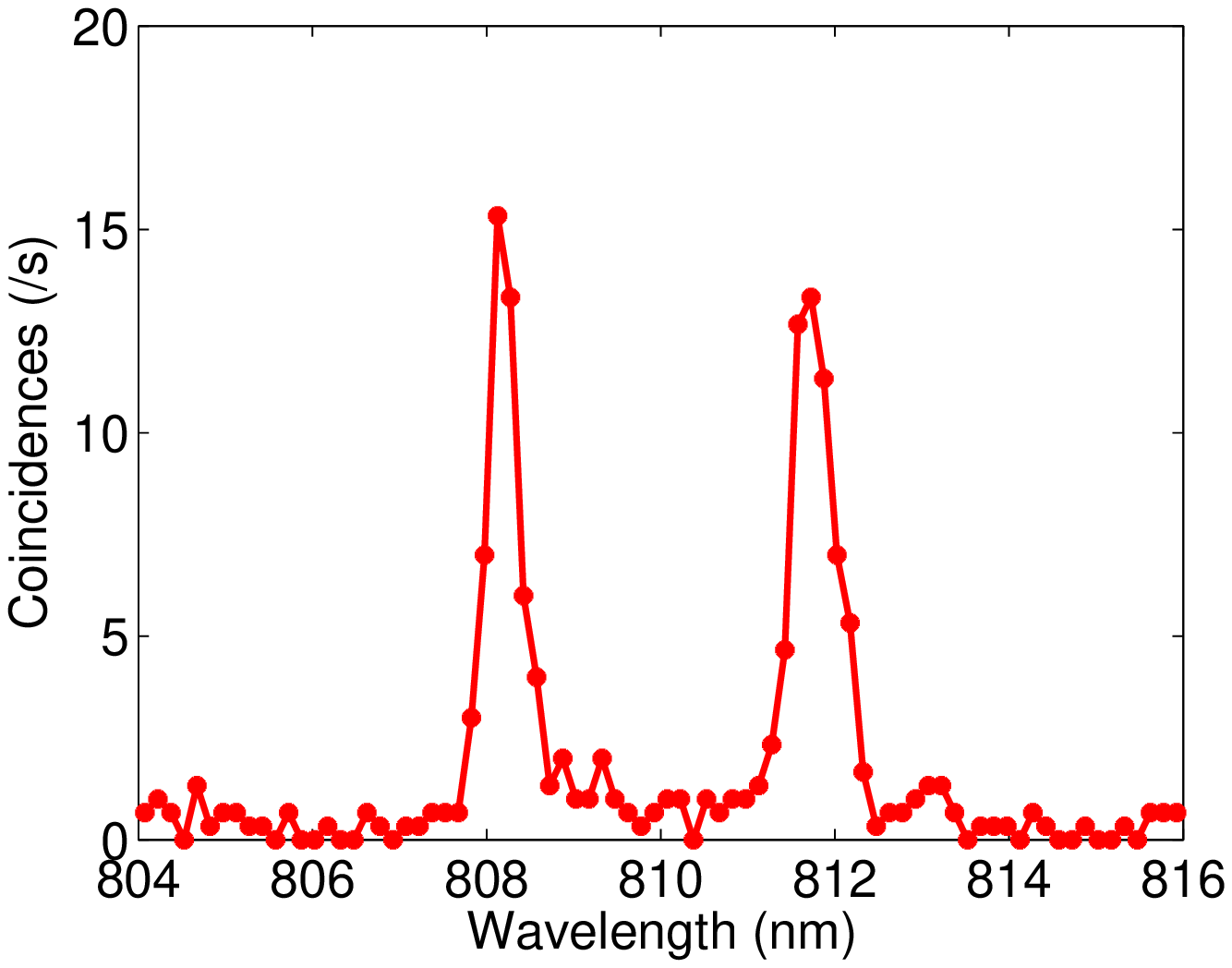}}
\subfigure[]{
\label{FigS3.sub.10}
\includegraphics[width=0.18\linewidth]{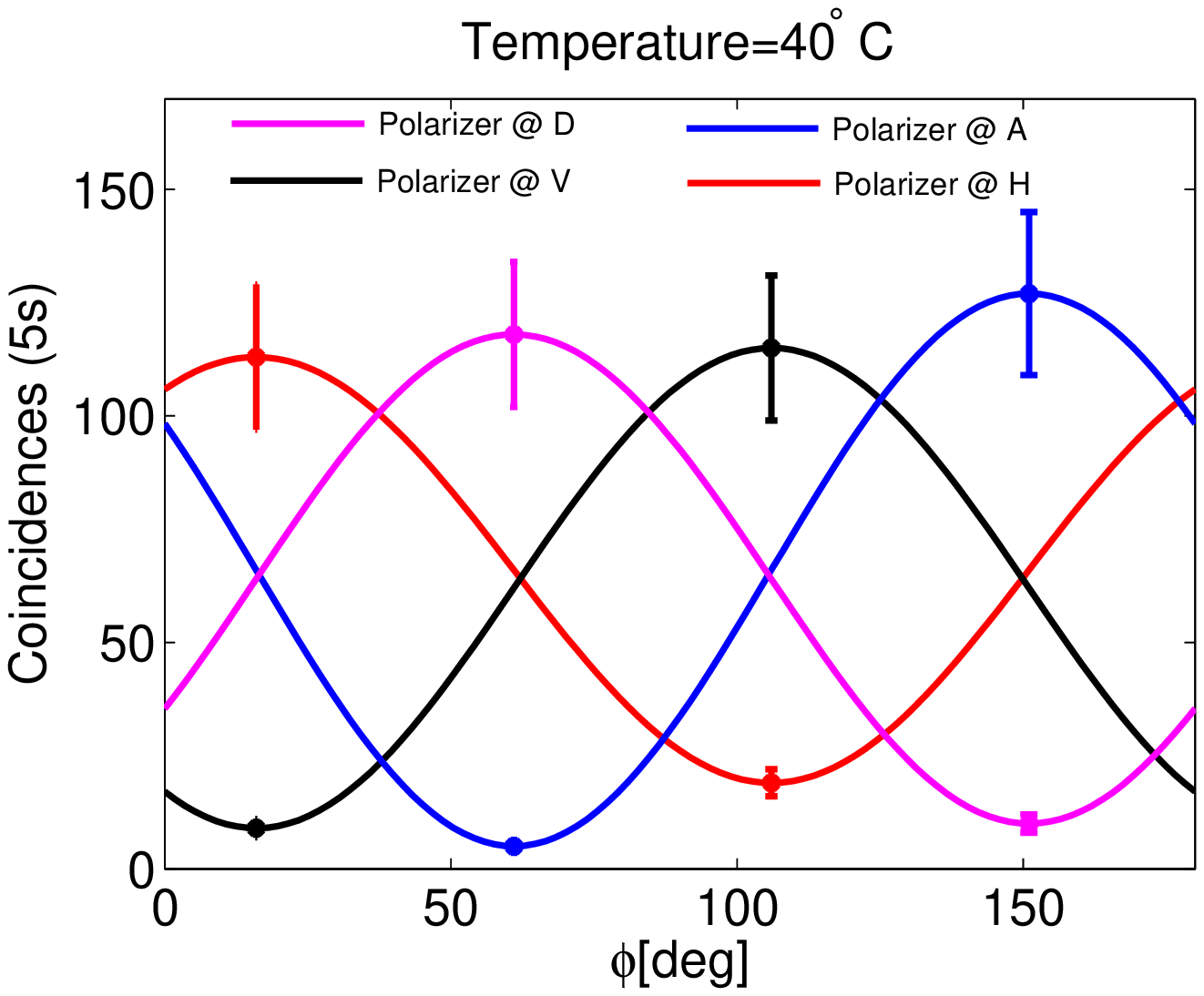}}
\caption{Verification of hyperentangled state after HOM interference. By setting relative path delay at zero, we observe (a) singles and (b) coincidences in opposite spatial modes, (c) singles and (d) coincidence in identical spatial modes and (e) characterization of polarization entanglement at highest interference peak position. By setting relative path delay at 0.08 mm, we observe (f) singles and (g) coincidences in opposite spatial modes, (h) singles and (i) coincidence in identical spatial modes and (j) characterization of polarization entanglement at lowest interference dip position.}
\label{figure_S3}
\end{figure*}
\indent In order to verify the quality of hyperentanglement after HOM interference for more practical quantum information applications, we build one monochrometer, consisting of two plane-convex lens and one reflective grating, to analyze the frequency correlation. Through experimental verification, this monochrometer reaches high resolutions up to 0.2 nm when rotating the grating by $0.01^\circ$ for each step. We first set the relative path delay at zero such that the interference fringe is at highest peak position. As shown in Fig.\ \ref{figure_S3}\textcolor{blue}{(a-d)}, we could observe two-fold coincidences of frequency bins in the opposite and identical spatial modes (assisted by in-fiber beam splitter). Almost all frequency coincidences are measured in opposite spatial modes. Figures \ref{figure_S3}\textcolor{blue}{(f-i)} demonstrate the two-photon coincidence envents in the scenario of setting the relative path delay at 0.08 mm such that the interference fringe is at lowest dip position. Now most of the coincidence events exist in identical spatial modes. Additionally, we observed polarization entanglement visibilities of $87\%$ in H/V basis and $82\%$ in A/D basis at maximum position (see Fig.\ \ref{figure_S3}\textcolor{blue}{(e))} and visibilities of $88\%$ in H/V basis and $78\%$ in A/D basis at dip position (see Fig.\ \ref{figure_S3}\textcolor{blue}{(j))} by measuring interference contrast. We attribute the decrease of polarization visibility to imperfect input hyperentanglement states, imperfect mode matching and residual misalignment at the PBS, finite PBS extinction ratio and accidental coincidences caused by interference.

\section{Fidelity lower bound for the discrete frequency subspace}\label{a:boundderivation}
In this section we demonstrate how to obtain a fidelity lower bound for the generated state $\rho$ in the discrete frequency subspace with respect to a maximally entangled state from the experimentally measured visibility
\begin{equation}\label{expvisibility}
 V_\omega := \frac{N_c^\text{max}-N_c^\text{min}}{N_c^\text{max}+N_c^\text{min}},
\end{equation}
where $N_c^\text{max(min)}$ is the maximum (minimum) number of coincidence counts as a function of the path delay (see Fig.\ref{Fig3.sub.2}).

We start by modelling the nonlocal measurement performed by the HOM interferometer and detection apparatus as a two outcome POVM, that acts on the joint frequency space of the two photons, corresponding to the bunching and the anti-bunching effects. The anti-bunching outcome corresponds to POVM element $\ketbra{\Psi^-_\omega}{\Psi^-_\omega}$, where $\ket{\Psi^-_\omega} = \frac{1}{\sqrt{2}} (\ket{\omega_1\omega_2} - \ket{\omega_2\omega_1})$, and the bunching outcome corresponds to POVM element $\mathbb{1}-\ketbra{\Psi^-_\omega}{\Psi^-_\omega}$, where $\mathbb{1}$ is the qubit identity operator. A coincidence detector count indicates an anti-bunching outcome while a single detector count indicates a bunching outcome.

The probabilty $p$ of an anti-bunching event is then given by
\begin{equation}
p = \text{Tr}(\ketbra{\Psi^-_\omega}{\Psi^-_\omega}\rho) = \frac{N_c}{N_\text{total}},
\end{equation}
where $N_c$ is the total number of coincidence counts and $N_\text{total}$ is the total number of detection events.

The path delay of one of the photons results in a phase shift $\phi$, which is modelled by the one-parameter qubit unitary $U_\phi$ given by
\begin{equation}
U_\phi=\left\{
\begin{matrix}
1&0\\
0&e^{i\phi}\\
\end{matrix}
\right\}.
\end{equation}

For a measurement of a pair of photons, one of which suffered a phase shift $\phi$, the probability of anti-bunching is then given by
\begin{equation}
p = \text{Tr}(\ketbra{\Psi^-_\omega}{\Psi^-_\omega}\,\rho_\phi)= \frac{N^\phi_c}{N_\text{total}},
\end{equation}
where we define $\rho_\phi:=(\mathbb{1}\otimes U_\phi) \rho (\mathbb{1}\otimes U^\dagger_\phi)$, $N^\phi_c$ is the number of coincidence counts at phase shift $\phi$ and the total number of detection events is constant for all $\phi$.

Now, notice that this probability corresponds exactly to the fidelity $F_\omega(\rho_\phi) = \text{Tr}(\ketbra{\Psi^-_\omega}{\Psi^-_\omega}\,\rho_\phi)$ of the state $\rho_\phi$ with respect to the maximally entangled state $\ket{\Psi^-_\omega}$, hence
\begin{equation}
F_\omega(\rho_\phi) = \frac{N^\phi_c}{N_\text{total}}.
\end{equation}

Setting $F^\text{max}_\omega:=\max_{\phi}F_\omega(\rho_\phi)$ and equivalently $F^\text{min}_\omega:=\min_{\phi}F_\omega(\rho_\phi)$ allows us to express the experimentally measured visibility $V_\omega$ in Eq. \ref{expvisibility} in terms of the fidelity:
\begin{equation}\label{expvisibility2}
 V_\omega = \frac{F^\text{max}_\omega-F^\text{min}_\omega}{F^\text{max}_\omega+F^\text{min}_\omega}.
\end{equation}

Calculating explicitly $F^\text{max}_\omega$ we get
\begin{align}
F^\text{max}_\omega =& \max_\phi \frac{1}{2} \Big[ \bra{\omega_1\omega_2}\rho\ket{\omega_1\omega_2} +\bra{\omega_2\omega_1}\rho\ket{\omega_2\omega_1} + \nonumber \\
&- e^{i\phi} \bra{\omega_1\omega_2}\rho\ket{\omega_2\omega_1} - e^{-i\phi} \bra{\omega_2\omega_1}\rho\ket{\omega_1\omega_2} \Big] \\
=& \frac{1}{2} \Big(\bra{\omega_1\omega_2}\rho\ket{\omega_1\omega_2} +\bra{\omega_2\omega_1}\rho\ket{\omega_2\omega_1}\Big) + \nonumber \\
&+ |\bra{\omega_1\omega_2}\rho\ket{\omega_2\omega_1}|, \label{Fmax}
\end{align}
and equivalently for $F^\text{min}_\omega$,
\begin{align}
F^\text{min}_\omega =& \frac{1}{2} \Big(\bra{\omega_1\omega_2}\rho\ket{\omega_1\omega_2} +\bra{\omega_2\omega_1}\rho\ket{\omega_2\omega_1}\Big) + \nonumber \\
&- |\bra{\omega_1\omega_2}\rho\ket{\omega_2\omega_1}|.
\end{align}

Substituting in $F^\text{max}_\omega$ and $F^\text{min}_\omega$ in  Eq. \ref{expvisibility2}, one arrives at
\begin{equation}
V_\omega = \frac{2|\bra{\omega_1\omega_2}\rho\ket{\omega_2\omega_1}|}{\bra{\omega_1\omega_2}\rho\ket{\omega_1\omega_2} +\bra{\omega_2\omega_1}\rho\ket{\omega_2\omega_1}}.
\end{equation}

To achieve the final goal of relating the measured quantity $V_\omega$ to the fidelity of the generated state $\rho$, we turn to the expression of $F^\text{max}_\omega$ in Eq. \ref{Fmax} and bound the first term of the sum according to
\begin{align}
\bra{\omega_1\omega_2}\rho\ket{\omega_1\omega_2} &+\bra{\omega_2\omega_1}\rho\ket{\omega_2\omega_1} \nonumber \\
&\geq 2 \sqrt{\bra{\omega_1\omega_2}\rho\ket{\omega_1\omega_2}\bra{\omega_2\omega_1}\rho\ket{\omega_2\omega_1}} \\
&\geq 2 |\bra{\omega_1\omega_2}\rho\ket{\omega_2\omega_1}|,
\end{align}
by applying the inequality $\sqrt{ab}\leq\frac{1}{2}(a+b)$ for non-negative real numbers $a$ and $b$ in the first step and the Cauchy-Schwarz inequality $|\bra{mn}\rho\ket{nm}|\leq\sqrt{\bra{mn}\rho\ket{mn}\bra{nm}\rho\ket{nm}}$ in the second step. Consequently,
\begin{align}
F^\text{max}_\omega &\geq 2 |\bra{\omega_1\omega_2}\rho\ket{\omega_2\omega_1}| \\
&= V_\omega \times \Big(\bra{\omega_1\omega_2}\rho\ket{\omega_1\omega_2}+\bra{\omega_2\omega_1}\rho\ket{\omega_2\omega_1}\Big).
\end{align}

Finally, we make the assumption of energy conservation, which implies $\bra{\omega_1\omega_1}\rho\ket{\omega_1\omega_1}=\bra{\omega_2\omega_2}\rho\ket{\omega_2\omega_2}=0$ and $\bra{\omega_1\omega_2}\rho\ket{\omega_1\omega_2}+\bra{\omega_2\omega_1}\rho\ket{\omega_2\omega_1}=1$, to arrive at our final fidelity lower bound of
\begin{equation}
F^\text{max}_\omega \geq V_\omega,
\end{equation}
concluding that the experimentally measured quantity $V_\omega$ is a direct lower bound for the fidelity of the prepared state in the frequency subspace with respect to a maximally entangled state, under the assumption of energy conservation.

Energy conservation is the only assumption about the source that is necessary to guarantee the validity of this lower bound. However, this is a very well physically motivated assumption that is also encouraged by the single photon spectra in Fig.~\ref{Fig3.sub.1}, which show two distinct frequency modes, symmetrically distributed around half the pump photon energy.

\end{document}